\documentclass{aa}
\usepackage[comma,authoryear]{natbib}
\usepackage{graphics}
\usepackage[latin1]{inputenc} 

\begin{document}

 \title{Using the Sun to estimate Earth-like planets detection capabilities.
  }
   \subtitle{II. Impact of plages. }


   \author{N. Meunier \inst{1}, M. Desort \inst{1}, A.-M. Lagrange \inst{1}  
	  }
   \authorrunning{Meunier et al.}

   \institute{
Laboratoire d'Astrophysique de Grenoble, Observatoire de Grenoble, Universit\'e Joseph Fourier, CNRS, UMR 5571, 38041 Grenoble Cedex 09, France\\
  \email{nadege.meunier@obs.ujf-grenoble.fr}
             }

\offprints{N. Meunier}

   \date{Received 23 October 2009 ; Accepted 10 December 2009}

\abstract{}
{Stellar activity produced by spots and plages affects the radial velocity (RV) signatures. Because even low activity stars would produce such a signal, it is crucial to determine how it influences our ability to detect small planetary signals such as those produced by Earth-mass planets in the habitable zone (HZ). In a recent paper, we investigated the impact of sunlike spots. We aim here to investigate the additional impact of plages.}
{We used the spot and plage properties over a solar cycle to derive the RV that would be observed if the Sun was seen edge-on. The RV signal comes from the photometric contribution of spots and plages and from the attenuation of the convective blueshift in plages. We analyzed the properties of the RV signal at different activity levels and compared it with commonly used activity indicators such as photometry and the Ca index. We also compared it with the signal that would be produced by an Earth-mass planet in the HZ. }
{We find that the photometric contributions of spots and plages to the RV signal partially balance each other out, so that the residual signal is comparable to the spot signal. However, the plage contribution due to the convective blueshift  attenuation dominates the total signal, with an amplitude over the solar cycle of about 8-10~m/s. Short-term variations are also significantly greater than the spot and plage photometric contribution. This contribution is very strongly correlated with the Ca index on the long term, which may be a way to distinguish between stellar activity and a planet.  }
{Providing a very good temporal sampling and signal-to-noise ratio, the photometric contribution of plages and spots should not prevent detection of Earth-mass planets in the HZ. However, the convection contribution makes such a direct detection impossible, unless its effect can be corrected for by methods that still need to be found. We show that it is possible to identify the convection  contribution if the sensitivity is good enough, for example, by using activity indicators. }

 \keywords{Techniques: radial velocities -- Stars: starspots -- 
Stars: planetary systems -- Stars: activity } 

\maketitle

\section{Introduction}

Among other techniques, the radial velocity (hereafter RV) technique has been very successful in detecting exoplanets. The increasing sensitivity of new instruments (for example, Espresso on the VLT or Codex on the E-ELT) is expected to give access to increasingly lower planet masses (down to Earth-like masses); however, it also makes the observed RV signal sensitive to other sources than planets, such as stellar activity and pulsations.

The influence of spots on the RV and bisector were first studied by \cite{saar97} and \cite{hatzes02}. This was followed by the precise study of the influence of a single spot on the signal \cite[][]{desort07} in various stellar conditions such as inclinations or angular rotation rates. In \cite{lagrange09}, hereafter Paper I, we simulated the RV signal due to spots that would be detected if the Sun was observed edge-on over a long period (typically one solar cycle) with different temporal samplings (between one day and 20 days) with only a few gaps. The advantage of such an approach compared to the previous ones is that we could take the full complexity of the activity pattern into account (structures, spatial distribution, temporal variations), because the signal from spots on various positions on the disk can partially balance  each other out. Our approach also allows us to study the temporal behavior of the signal (such as with periodograms, which are also very useful tools to detect planets in the RV signal). We derived a typical RV signal of 0.45~m/s (rms), with maximum peaks at $\pm$2~m/s, especially during the solar maximum (R'$_{HK}$ $\sim$-4.85). During the solar minimum (R'$_{HK}$ $\sim$-5.05), the rms falls to 0.16~m/s with maximum peaks at $\pm$0.6~m/s. Then we estimated the impact of these spots on the detection of Earth-like planets and showed that a very good temporal sampling over a long period was mandatory for detecting Earth-mass planets in the habitable zone, depending on the cycle phase.

However, spots are not the only source of RV variations. The presence of bright plages also induces a variable signal in RV, as for spots. They are well-known on the Sun, but there are indirect indications that they are also present on other stars, as they sometimes dominate the photometry \cite[e.g.][]{lockwood07}. Their contrast in temperature is less than for spots, but their area is much larger. Their contribution is therefore expected to be significant. Such a contribution to the RV has never been studied in precise detail. Because they are bright, we expect plages located at the same position as spots to produce an RV signal anti-correlated with the one from the spot (see Paper I). However, because they are not localized exactly at the same position (in addition to their different spatial extension) and have a position-dependent contrast, and because in practice there is no one-to-one correspondence between the two types of structures \cite[the plage to spot area ratio presents a very large dispersion,][]{chapman01}, the residual signal will be significant. In addition, there are many small plage-like structures outside active regions, called the network, contributing significantly to the photometric variations of the Sun. It is crucial to study the two contributions (spots and plages) separately and in detail because the respective contributions of plages and spots may be different on other stars \cite[][]{lockwood07}.

Another contribution of plages comes from the attenuation of the convective blueshift \cite[due to the presence of granulation, e.g.][]{gray09} when magnetic field is present. Rough estimates have been made on individual lines \cite[][]{saar97,saar03,kurster03} but the impact on the RV variations derived from the cross-correlation function (CCF) has not been studied yet. 

The objective of this paper is therefore to estimate the contribution of plages to the RV signal (either from the photometry or the convection), to compare it to the signal produced by spots at the same time, and to study the resulting signal. As in Paper I, we consider a Sun seen edge-on over a solar cycle. In Sect.~2, we describe in detail the data set (spots and plages), the data processing, and the observables we obtain. Special care is given to the comparison with the observed photometry (total solar irradiance) in order to validate the parameters used to make the simulation. The resulting RV are analyzed in Sect.~3 and compared with other observables (photometry, Ca index) in Sect.~4. In Sect.~5, we study the influence of this signal on Earth-like planet detection, and conclude in Sect.~6.

\section{Description of the simulations}

In this section we first describe the input data: the list of sunspots and plages that is used as input to our simulation tool; the way we derive the temperature contrasts for these structures; the influence of convective inhomogeneities. Then we briefly describe our simulation and the produced observables, which are similar to what was done in Paper I.

\subsection{Input data}


To get a good temporal sampling of plages, we used MDI/SOHO magnetograms \cite[][]{Smdi95} between May 5, 1996 (julian day 2450209) and October 7, 2007 (julian day 2454380). This covers a solar cycle with a temporal sampling of about 1 day. We therefore used a different sunspot data set from Paper I to cover the full plage data set, although there is a significant overlap between julian days 2450209 and 2453004. We chose the sunspot group data provided by USAF/NOAA over the same period (http://www.ngdc.noaa.gov/stp/SOLAR/). 

This sunspot data set has a few gaps. Some of them correspond to days without any spot while others are true gaps in the data. We removed the latter from the MDI data set by checking the Wolf number. In the following, there are therefore plage structures every day, but in some cases there may be no spot on the Sun surface. This leads to 3586 days, for a total duration of 4171 days (i.e. a 86\% coverage, which is similar to what we had for spots in Paper I). 


\begin{figure*} 
\includegraphics{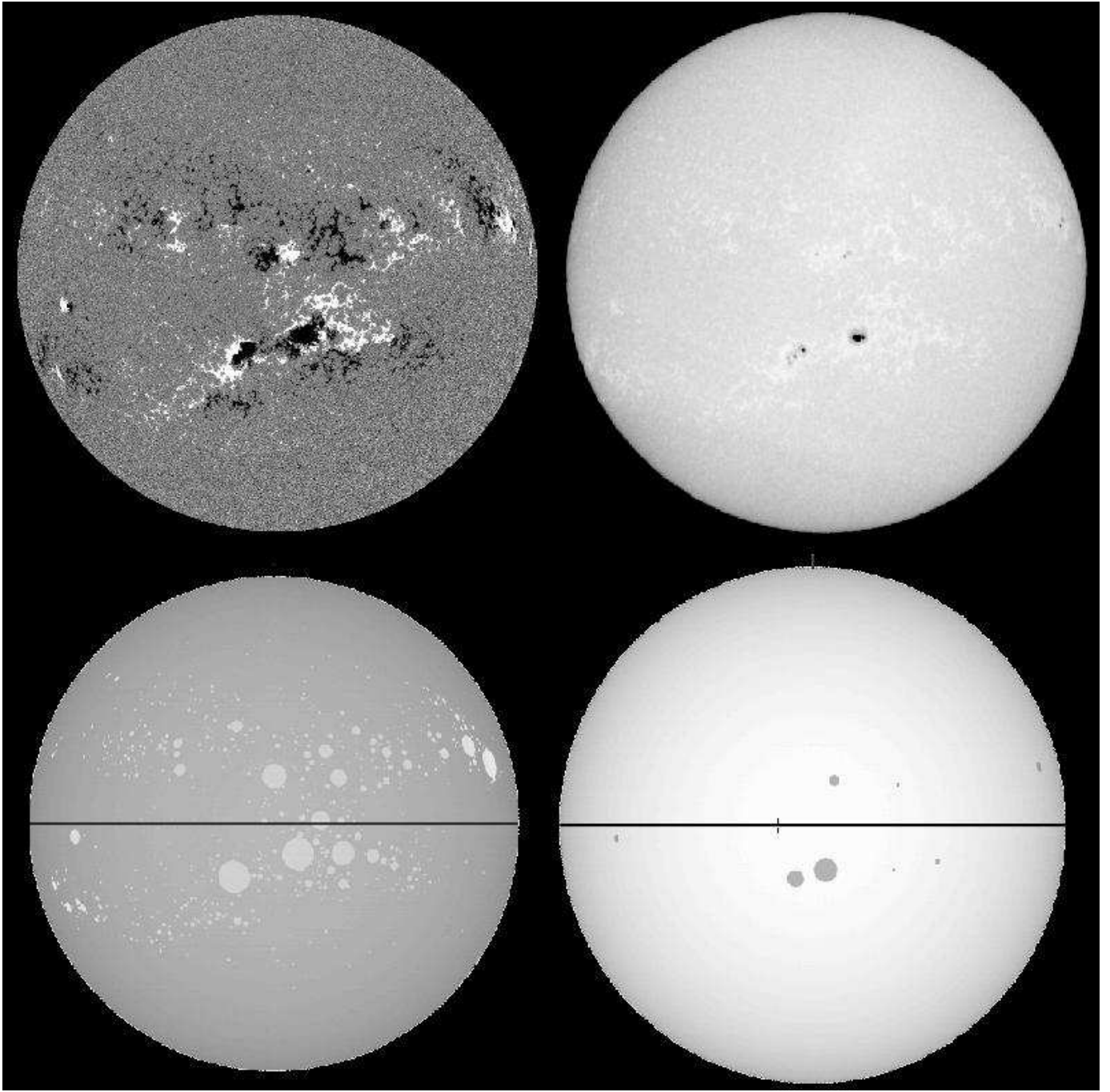}
\caption{{\it Upper panel}: MDI magnetogram for September 14, 2002 at 4:48 UT ({\it left}) and Meudon photospheric spectroheliogram at 6:41 UT ({\it right}). {\it Lower panel}: Simulated map showing the plages derived from the magnetogram with a contrast increased by a factor 10 for clarity ({\it left}) and the spots from the USAF/NOAA data set at 12:40 UT ({\it right}).}
\label{imag}
\end{figure*}

\begin{figure} 
\vspace{0cm}  
\includegraphics{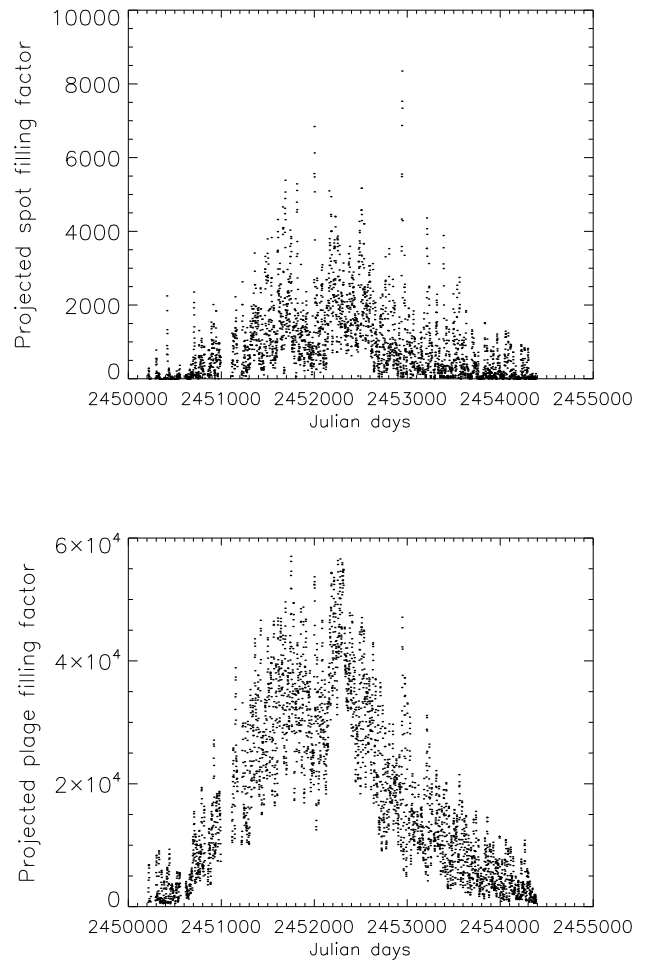}
\caption{{\it Upper panel}: projected spot filling factor versus time, in ppm of the solar disk. {\it Lower panel}: same for plages.  }
\label{ff}
\end{figure}

The spot data are used without any further processing and provide 35207 sunspot groups, with the smallest size 10~ppm (of the solar hemisphere). This size corresponds to a radius of 3.1~Mm, assuming a circular shape. We note that the surface coverage of the Debrecen data used in Paper I is larger by a factor of about 1.57 when comparing the same days. The impact of this difference is investigated in the next section. 

Plages (in active regions) and network structures (bright magnetic structures outside active regions) are extracted from MDI magnetograms by applying a threshold of 100~G for $\theta$ up to 72$^{\circ}$ (with $\theta$ the angle between the line-of-sight and the normal to the solar surface) in order to avoid the noise close to the limb. This threshold is quite conservative for limiting the influence of the noise (about 20~G), and is slightly larger than the one used by \cite{fligge00} on the same data. For the same purpose, the smallest structures are eliminated. The smallest size is then 3~ppm (of the solar hemisphere), which corresponds to a radius of 1.7~Mm assuming a circular shape. This is smaller than the smallest spots we consider because we also include network structures and not only plages from active regions. This leads to 1803344 structures, from the quiet network to large plages in active regions. The relative contribution of plages and network structures is discussed in Sect.~2.3.2. 

The structures extracted from the magnetograms using this simple segmentation also include spots of course. With the present context, this is not critical, however, since the surface coverage of spots is much smaller. Typically, the filling factor of spots (determined from the USAF data above) represents about 6\% of the filling factor of structures determined from the magnetograms. 
Furthermore, modifications of the threshold used to identify plages lead to modification of that filling factor by about 10\% in the case of large structures and up to 20\% for small network features, without changing their visual identification significantly. This gives an idea of the uncertainty on the filling factor. The influence of this variation is also investigated in the next section. 

The first input to our simulation tool (for both spots and plages) then consists in the list of structures: size in ppm of the solar hemisphere (before correction for projection effects), latitude, and longitude.
The projected filling factors are shown in Fig.~\ref{ff}. The correlation between the spot and plage filling factors is 0.77. This reflects the correlation between the two, both on long (cycle) and short time scales (rotation), as active regions usually include spots, but also the large dispersion of the plage-to-spot size ratio and the inclusion of the network structures (not spatially correlated with spots as they are located outside active regions). The plage filling factor is about one order of magnitude more than the spot filling factor.

\subsection{Contrast choices}

\subsubsection{Spot and plage observed contrasts}

There is a large dispersion in the literature concerning the spot and plage contrasts. There is also a large temperature difference from one structure to the next. 
It is therefore difficult  to directly use the published contrasts found in the literature \cite[for example][]{frazier71,lawrence88,ermolli07}, because they are very sensitive to the way plages are defined. The large dispersion between individual plages, as well as the differences between plages and network structures \cite[][]{foukal91,worden98,ortiz02,ortiz06} also adds to the uncertainty of these measurements. 
To realize a consistent simulation, we computed the photometric contribution of each kind of structure (spots, plages, network) and compared it with the observed total solar irradiance (hereafter TSI).   
In the following we use the TSI compiled by Claus Fr\"ohlich and Judith Lean (http://www.ngdc.noaa.gov/stp/SOLAR/) for the 1996--2003 period and described in \cite{frohlich98}. This comparison is therefore made over 2263 points.  
 
In Paper I, we used a spot temperature deficit of 550~K, as this lies well within the range of observations and simulations \cite[e.g.][]{chapman94,fligge00, krivova03}. The situation is more complex for plages. We can either use a shape such as the one used by \cite{lanza04,lanza07} and in Paper I, i.e. a temperature excess as a function of 1-$\mu$ (where $\mu=\cos\theta$). We can also use a slightly more complex function such as in \cite{unruh99} and used by \cite{fligge98,fligge00} to reconstruct the solar irradiance, to account for the difference from zero contrast at disk center. In the following, we use a second-degre polynomial starting with a shape similar to the one modeled by \cite{unruh99}, as it gives a slightly better $\chi^2$ when comparing the simulated irradiance to the observed one (see next section).

\subsubsection{Determination of the spot and plage contrast}

\begin{figure} 
\vspace{0cm}  
\includegraphics{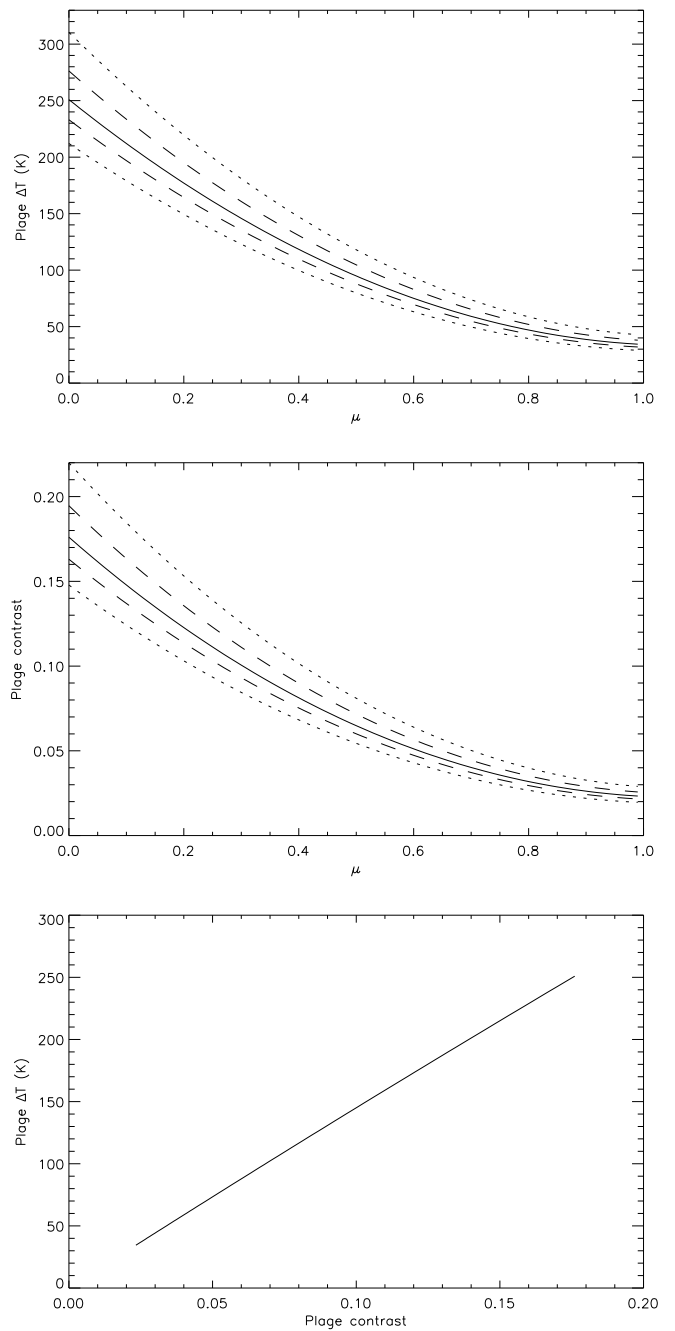}
\caption{{\it Upper panel}: Plage temperature excess versus $\mu$ (solid line), and with a modification of the plage size of $\pm$10\% (dashed lines) and $\pm$20\% (dotted lines), see text for details. {\it Middle panel}: Same for the contrast at 600~nm. {\it Bottom panel}: Plage temperature excess versus the contrast.}
\label{dtplage}
\end{figure}

\begin{figure} 
\vspace{0cm}  
\includegraphics{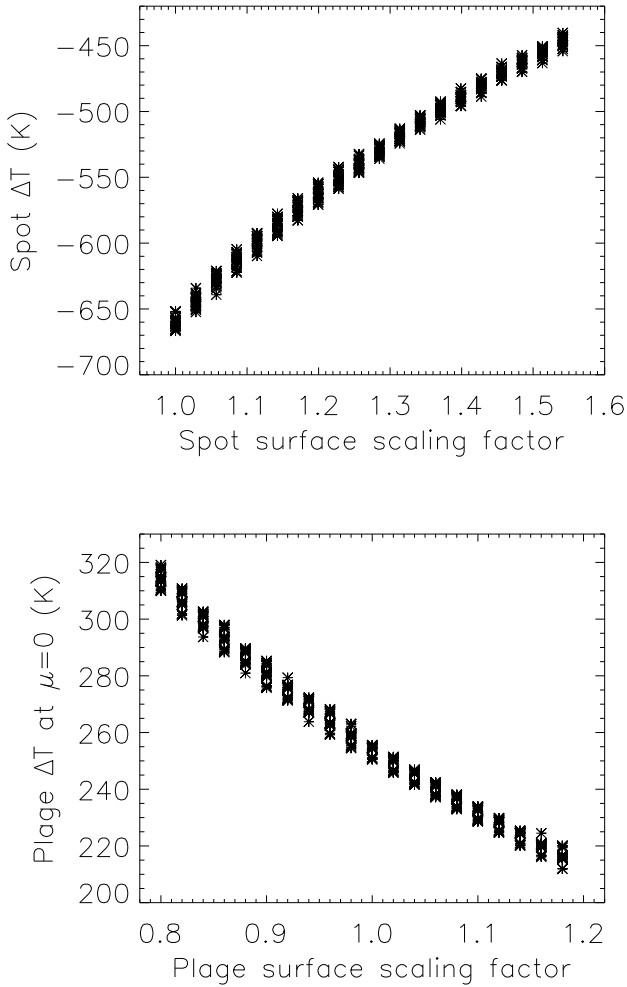}
\caption{{\it Upper panel}: Spot temperature deficit versus the spot-surface scaling factor for the explored range of scaling factors (see text for details). For a given spot-surface scaling factor, the various points correspond to different plage-surface scaling factors. {\it Bottom panel}: Same for plages at $\mu$=0.}
\label{t_surf}
\end{figure}

We used the following procedure to determine the spot and plage contrasts. 
We first consider a spot temperature deficit\footnote{with respect to the photospheric temperature of 5800~K} $\Delta T_s$=-550~K and a plage contrast\footnote{defined as $(S_{pl}-S_{ph})/S_{ph}$, with $S_{ph}$ the photospheric irradiance and $S_{pl}$ the plage irradiance} $C_p$=0.339-0.563$\mu$ +0.270$\mu^2$~K, which corresponds to the contrast provided by \cite{unruh99} for wavelengths in the range 470-550~nm. 
We then computed a contrast at 600~nm assuming a Planck law, which leads to estimate the contribution of spots and plages to the TSI. We chose 600~nm for reference because that is where the contrast is close to the average contrast over 350-1000~nm, representative of the bolometric value \cite[][]{gondoin08}, hence the TSI. 

This leads to two series, $I_{sp}$ and $I_{pl}$, respectively representing the contribution of dark and bright features to the solar irradiance, in fractions of the TSI of the quiet Sun (i.e. contrasts). When multiplied by the quiet Sun irradiance $S_{ref}$ they allow a direct comparison with the observed TSI $S_{obs}$ (in W/m$^2$) by computing 

\[ S_{mod}=S_{ref} (I_{sp}+I_{pl}+1), \] 

\noindent which sums the contributions of spots, plages (including the network), and quiet Sun respectively.  
We can adjust $S_{mod}$ to $S_{obs}$ by applying a scaling factor to each contribution and fitting them to minimize $\chi^2=\sum (S_{mod}-S_{obs})^2/(N-N_{par}) $ where 

\[ S_{mod}= S_{ref} f_{ref} (f_{sp} I_{sp}+f_{pl} I_{pl}+1),\] 

\noindent and $N$ the number of points, $N_{par}$ the number of parameters.
The parameter $f_{ref}$ allows adjustment of the quiet-Sun level, while $f_{sp}$ and $f_{pl}$ allow the contrast to be varied for spots and plages respectively. 
For our input spot and plage data as described above, we obtain the following best choice: $S_{ref}$= 1365.46 W/m$^2$ \cite[which is very close to the one given by][for the cycle minimum in 1996]{frohlich09}, $\Delta T_s$=-663~K and $\Delta T_p$=250.9-407.7$\mu$+190.9$\mu^2$~K. Figure~\ref{dtplage} (solid lines) shows the plage excess temperature as a function of $\mu$, as well as the corresponding contrast at 600~nm. This contrast is compatible with the literature. The correlation between the simulated and observed irradiance is 0.89.  

\begin{figure} 
\vspace{0cm}  
\includegraphics{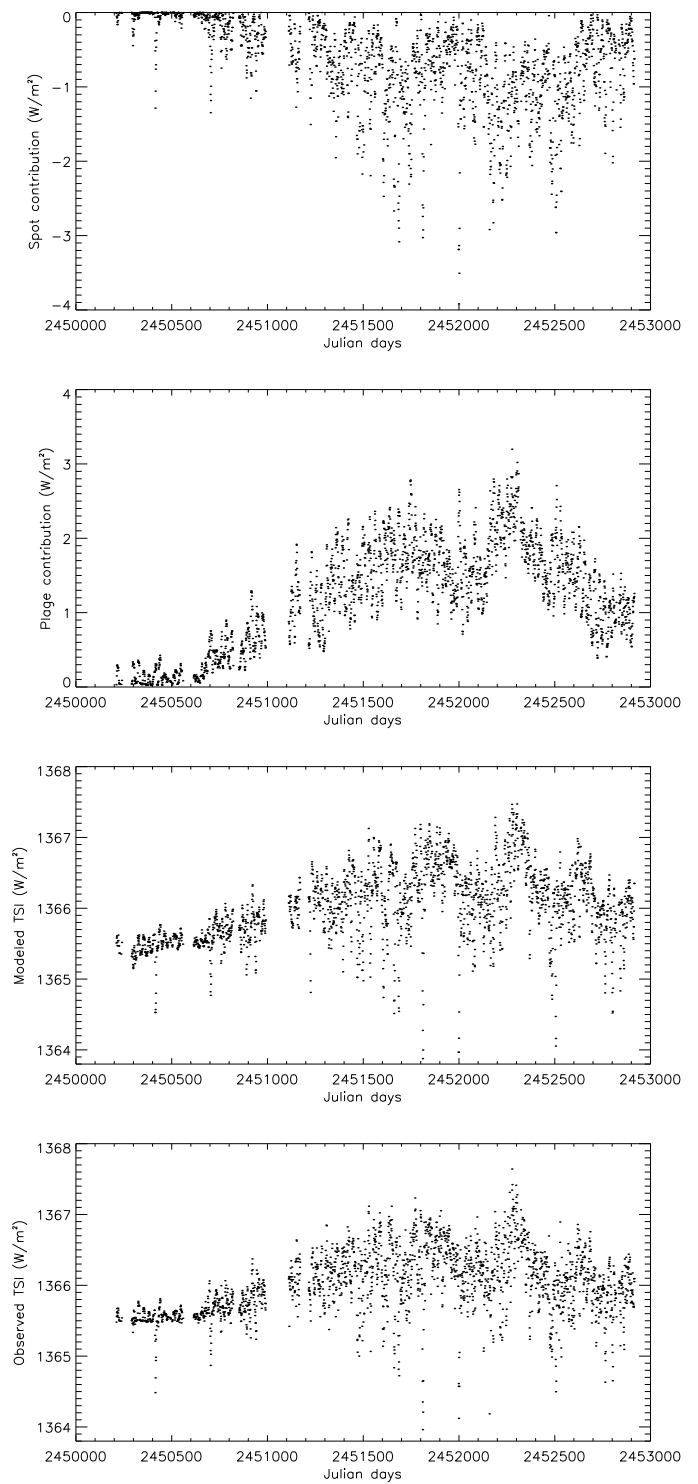}
\caption{{\it First panel}: Reconstructed spot contribution to irradiance. {\it Second panel}: Reconstructed plage contribution to irradiance. {\it Third panel}: Total reconstructed irradiance. {\it Fourth panel}: Observed irradiance.}
\label{phot}
\end{figure}

The final reconstructed irradiance versus the observed TSI will be shown in Sect.~2.5, as here we have made a simplified computation to correct for projection effects, we attribute to each structure the value of $\mu$ corresponding to the center of the structure, while in the following we build maps of the solar surface. We also neglect the center-to-limb darkening (which will be taken considered in Sect.~2.5 where we make a more precise computation of $S_{mod}$). Figure~\ref{phot} shows the contribution of spots and plages to the irradiance, as well as the simulated irradiance used here to derive the set of parameters. Our results are quite similar to those of \cite{frohlich98}, and we reproduce the observed irradiance.

\subsubsection{Uncertainties}

To study the influence of our input data set (USAF sunspot group and plage extraction from MDI magnetograms), we explored the parameter space related to the structure sizes by applying a certain scaling factor to them, which should represent the uncertainty we have on the structure sizes. For spots, we applied scaling factors between 1 (original USAF values) and 1.57 (corresponding the Debrecen filling factor), i.e. we allow the spot size to vary by about 50\%. The smaller filling factor in the USAF catalog may stem from missing very small spots \cite[][]{balmaceda09}, at least partially. The correction also included a small dependence on the size, as we have been able to compare specific regions, which shows that, as expected, the correction is relatively large for small structures. For plages, we applied scaling factors between 0.8 and 1.2, which takes the assumption concerning the inclusion of spots into account, as well as a margin for the choice of the threshold. We therefore allowed the plage and network sizes to vary by $\pm$20\%.

For each new series of spot and plage sizes, we searched for the contrasts that provide the best fit to the observed TSI, following the method described in Sect.~2.2.2. 
Indeed, at first order, if one slightly overestimates the plage areas, for example, it can be compensated for a larger temperature excess. This is true for the reconstitution of the solar irradiance and also for RV computations as it also is a photometric effect.  
The quiet-Sun solar irradiance  is fixed to our best value above, as it is very robust. 

The results are as follows. The corresponding $\Delta T_s$ ranges between -670~K (for a spot scaling factor of 1) and -450~K (for a spot scaling factor of 1.57), with a strong correlation as expected (see Fig.~\ref{t_surf}). At $\mu$=0, $\Delta T_p$ varies between $\sim$315~K (for a plage scaling factor of 0.8) and $\sim$215~K (for a plage scaling factor of 1.2), and between $\sim$43~K and $\sim$29~K at $\mu$=1, here again with a strong anticorrelation because the areas and temperature excess compensate for each other. This anticorrelation is shown in Fig.~\ref{t_surf} for $\mu$=0. Figure~\ref{dtplage} also shows the plage-temperature excess (and contrast at 600~nm) versus $\mu$ for the whole domain.  
Over the whole domain, the $\chi^2$ remains very low. It does not show any systematic effect when changing the plage surface. The $\chi^2$ is, however, smaller for the spot scaling factor of 1 compared to 1.57. It is also very close to the minimum as a test of a few values below 1 for the spot scaling factor shows that the $\chi^2$ then increases again. 

Finally, we note that in Paper I we used the Debrecen data set, which corresponds to the spot-surface scaling factor of $\sim$1.6 in Fig.~\ref{t_surf}. We used a spot temperature deficit of -550~K, which lies in the middle of our range, instead of -450~K. We therefore expect our spot RV signal to be weaker by about 20\% than that of Paper I, in which the spot temperature deficit could not be validated precisely using the observed irradiance because plage were not taken into account. 

We are therefore confident that our choice of temperature contrasts will provide the correct amplitude of the RV signal, as we are including two crucial ingredients in our simulation: a correct amplitude due to a good combination between  area and contrast, as checked by the comparison with the observed irradiance, and a realistic temporal evolution of structures thanks to the temporal sampling and coverage. 
 
\subsection{Convective inhomogeneities}

\subsubsection{Estimation of convective shifts}

In addition to the photometric effect described above, another effect can lead to RV and bisector variations. The presence of convection in the photosphere produces a blueshift of spectral lines, as well as a distortion of these lines \cite[e.g.][]{dravins81,dravins82,dravins99,Liv99}. This comes from the correlation between velocities and brightnesses in granules. Granules are convective cells at the 1~Mm scale, in which upward motions associated to bright area occupy a larger surface than downward motion area (dark lanes), leading to blueshifts and to line distortion. However, where magnetic field is present, it is well known that convection is affected, and is greatly reduced \cite[e.g.][]{HNM91,TTT87,SGS88}. This strongly impacts the lines by locally  modifying their shape and producing a redshift that compensates for the convective blueshift \cite[][]{cavallini85,immer89,BS90,HM90,guenther91,HNM94}. This influence of the magnetic field on the lines has also been modelled by \cite{marquez96b}; therefore, we do expect this blueshift to be modified by the presence of plages (as well as spots, but their contribution is small as they represent a much smaller surface and are darker).

In the following, we estimate the amplitude of the convective blueshift in the quiet Sun.  
The convective blueshift largely depends on the line depth, and a large dispersion is present in the literature as most studies focus on a small number of lines. We have used the detailed results of \cite{gray09} for the Sun and derived a blueshift as a function of the line depths using a linear function. Then, on the solar spectrum obtained by Delbouille, Neven, and Roland (retrieved from the BASS2000 data base), we have identified the lines in our spectral range of interest, retrieved their position and depth, then simulated a spectrum in which each line has the same depth but is shifted by a different amount according to its depth and following \cite{gray09}. We used a linear relation between the line depth and the blueshift, which is extrapolated to lines outside the range of depth considered by \cite{gray09}. Finally, we computed the shift that would result from a cross-correlation applied on such a spectrum. We therefore consider the distribution of line depths in the spectrum and that cross-correlation functions (used later to derive the RV) are more sensitive to deep lines than weak lines. This produces a convective blueshift corresponding to the integrated light of about 200~m/s that is representative of the whole spectrum. The local (vertical) convective blueshift is therefore 285~m/s (the corresponding factor being the result of projection effects combined to the center-to-limb darkening). 
It is similar to the value used by \cite{kurster03} derived from \cite{dravins99}, but higher than the value considered by \cite{saar03}, about 90~m/s for the $v$sin$i$ we consider. 

\cite{gray09} studied the blueshift only for a quiet Sun (or star). On the other hand, \cite{BS90} studied the line shape and shifts for different magnetic filling factors (including the quiet Sun), by observing at different positions in a plage. 
They find a blueshift in plages which is 1/3 of the value in the quiet Sun. We therefore use in the following an attenuation of the convective blueshitf of 2/3 of our values, i.e. 190~m/s. The results of \cite{cavallini85} are also compatible with this result. We expect the attenuation to be around zero in spot umbra \cite[][]{solanki86, martinez97}, but the contribution of the umbra is very small. 
Given that the spread in convective blueshift associated to plages in the literature, about 50~m/s, we consider that, in the presence of plages and network (see next section), the convective blueshift is attenuated down to 190~m/s, with an uncertainty of $\pm$50~m/s. We neglect in this work the contribution of the Evershed effect (mostly associated to spot penumbra) and supergranulation, because we expect them to only increase slightly the jitter on scales from days to weeks.

\subsubsection{Notes on the convective shifts}

All solar observations of the line distortions and shifts due to magnetic fields have focused on plages. Our data set includes a significant contribution from the network. We also know that some abnormal granulation is associated to it, for example, to network bright points \cite[e.g.][]{muller89}. More recently, \cite{morinaga08} have shown the suppression of convection around small magnetic concentrations; however, there is no specific study of the lineshift associated to the network that we are aware of. In the following, we consider the RV signal associated to all structures, but will also estimate it for regions of a signifiant size only in order to test the possible range of variation. There is of course a continuum in size from the largest active regions to the small network structures \cite[][]{Meunier03}, but only considering sizes above 100~ppm (corresponding to a radius of 10~Mm) would probably  give a good estimation of a lower limit for the convection contribution to the RV. This is discussed in Sect.~3.3.2. 

Finally, it should be noted that the final RV signal due to convective inhomogeneities is sensitive to the actual areas of structures and to the $\Delta V$ we consider. As before, for area and temperatures, there is a trade-off between the two. However, unlike the temperature, we do not have the possibility of testing it independently.

\subsection{Simulated spectra}

Simulated spectra are computed as described in Paper I. A rigid rotation of 1.9~km/s at the equator is assumed. For each type of structure (spots and plages), we build a 3D hemisphere map of the visible Sun seen edge-on. We assumed circular shapes both cases, mostly for simplicity as we do not expect the shape to significantly impact on our result, and more importantly to be able to extrapolate easily to other stellar cases in future works. For the photometric contribution of magnetic structures, the resulting spectrum is then computed assuming that spots (respectively plages) emit a solar-like spectrum following a black body law with a temperature as described in the previous section. To take the convection into account, we consider a solar-like spectrum (black body with the photospheric temperature) shifted by a velocity perpendicular to the solar surface (which produces a redshift) of 190~m/s.

\subsection{Computation of the observables}

\begin{figure} 
\vspace{0cm}  
\includegraphics{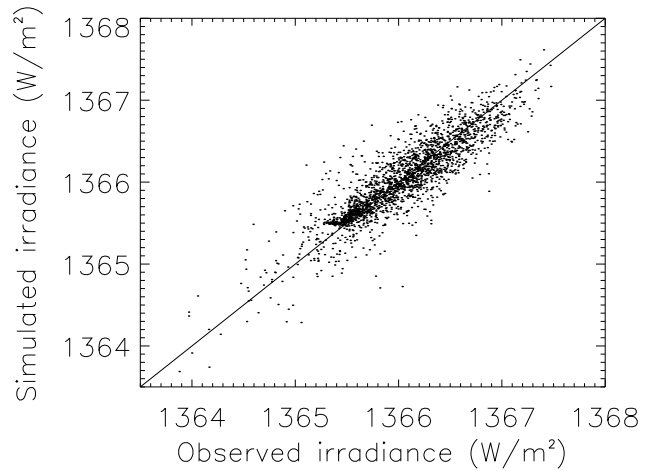}
\caption{Total reconstructed irradiance versus the observed irradiance (W/m$^2$). }
\label{phot2}
\end{figure}

The observables are then computed using the SAFIR software \cite[][]{galland05} as on actual stellar spectra. In this paper we focus on the analysis of the RV. As in Paper I, computations were made only on one spectral order (order \#31) as the whole process is very time consuming. 
We therefore obtained three series of RV, due to spots, to plages (and network) through their photometric contribution (hereafter the plage signal), and to the influence of plages on convection (hereafter the convection signal). The bisector velocity span (hereafter BVS) due to convective inhomogeneities has a very complex influence on the final line shape, and is not characterized with sufficient precision in the literature. We therefore only analyze the RV in this paper.

We also computed the photometry at 600 nm to check that we obtain the expected agreement. Figure~\ref{phot2} shows  the good correlation (0.89) between the final simulated photometry and the observed TSI. The residual difference between the model and the simulation largely comes from the strong temperature variation from one structure to the next. However, we capture the essential variability of the photometry.

\subsection{Simulation of an active star with a planet}

To test the planet detectability providing the activity we model, we added the signal of a 1 M$_{\rm earth}$  planet located at 1.2 AU (see Paper I) to the activity signal, as well as some random noise at various level. This is analyzed in Sect.~5.

\section{Results}

In this section we present the different contributions to the RV signals over the whole period. In addition, in order to have a better idea of the dependence on the activity level, we also show the results for two 9-month periods as in Paper I: a period of low activity from July 1, 1996 (JD 2450266) to April 1, 1997 (JD 2450540), leading to 213 data points over 274 days, and a period of high activity from February 1, 2000 (JD 2451576) to November 1, 2000 (JD 2451850), leading to 274 data points, also over 274 days. They will be denoted as $``$Low" and $``$High", respectively, in the next figures and tables.  

In addition to the total RV and the 3 individual components, we also discuss the contributions due to spots and plages separately (the total being denominated as the spot+plage signal) and compare them with the convection signal. This is important as they have a different origin (even if the convection signal relates to plages): on a star with similar plages but much weaker convection, the plage signal would remain the same but the convection signal would become very small. 

\subsection{RV variations}

\begin{figure*} 
\vspace{0cm}  
\includegraphics{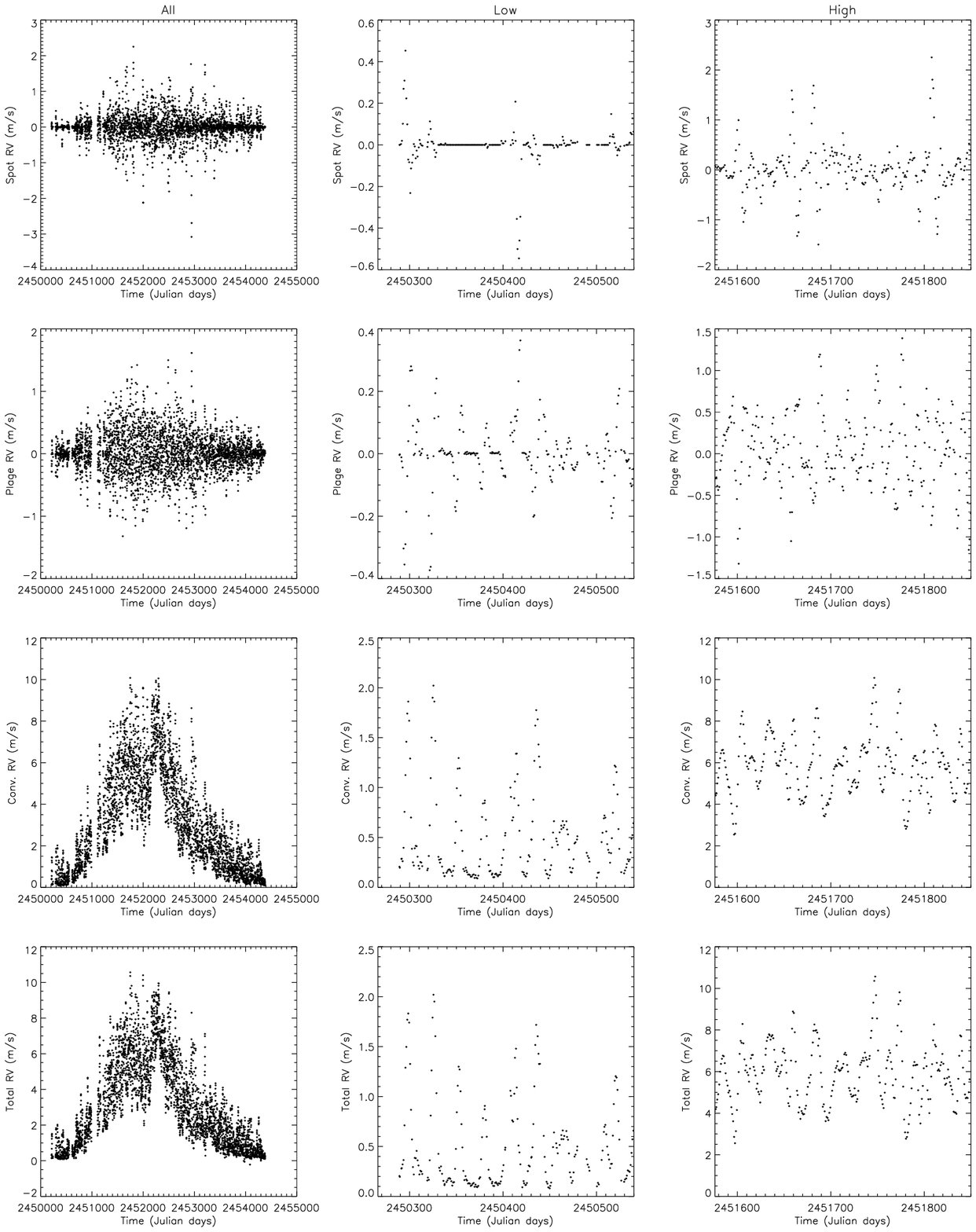}
\caption{{\it Left column}: RV (m/s) for spot signal, plage signal, convection signal, and total ({\it From top to bottom}), for the whole data set. {\it Middle column}: Same for the low activity period. {\it Right column}: Same for the high activity period.}
\label{var_rv}
\end{figure*}

\begin{figure} 
\vspace{0cm}  
\includegraphics{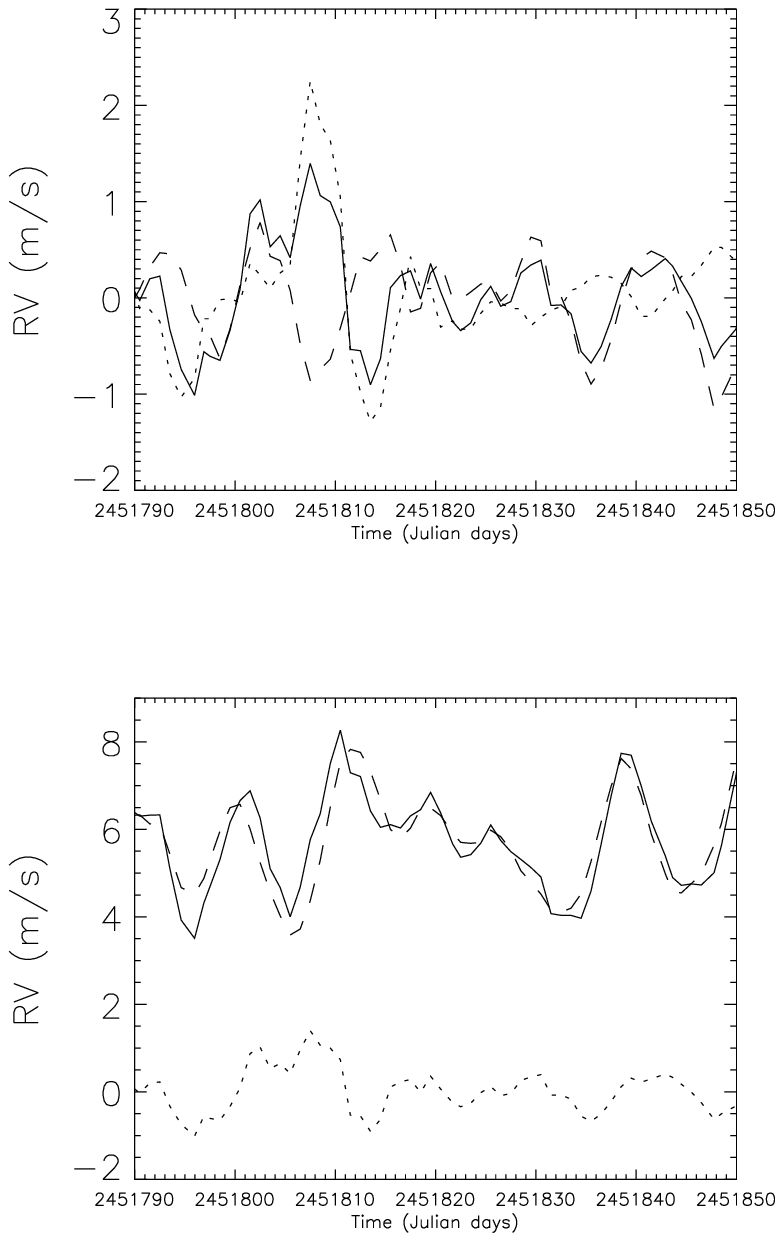}
\caption{{\it Upper panel}: RV signal from spots (dotted line), plages (dashed line), and spots+plages (solid line) in m/s, for julian days 2451790--2451850, i.e. during the high activity period. {\it Lower panel}: Same for spots+plages (dotted line), convection (dashed line), and total (solid line).}
\label{varloc}
\end{figure}

\begin{figure} 
\vspace{0cm}  
\includegraphics{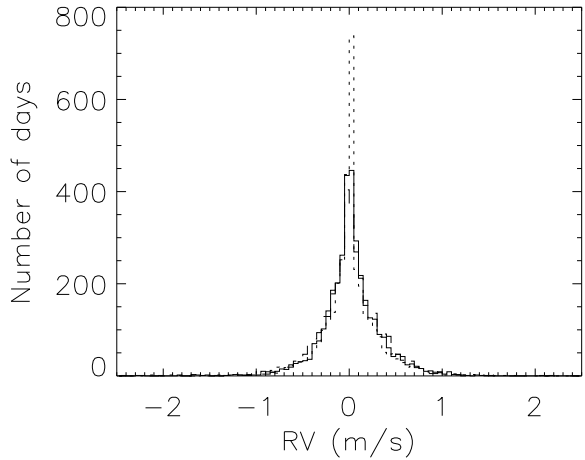}
\caption{Histogram of RV (m/s) for spots+plages (solid line), spots (dotted line), and plages (dashed line), for the whole period.}
\label{histo}
\end{figure}

\begin{figure} 
\vspace{0cm}  
\includegraphics{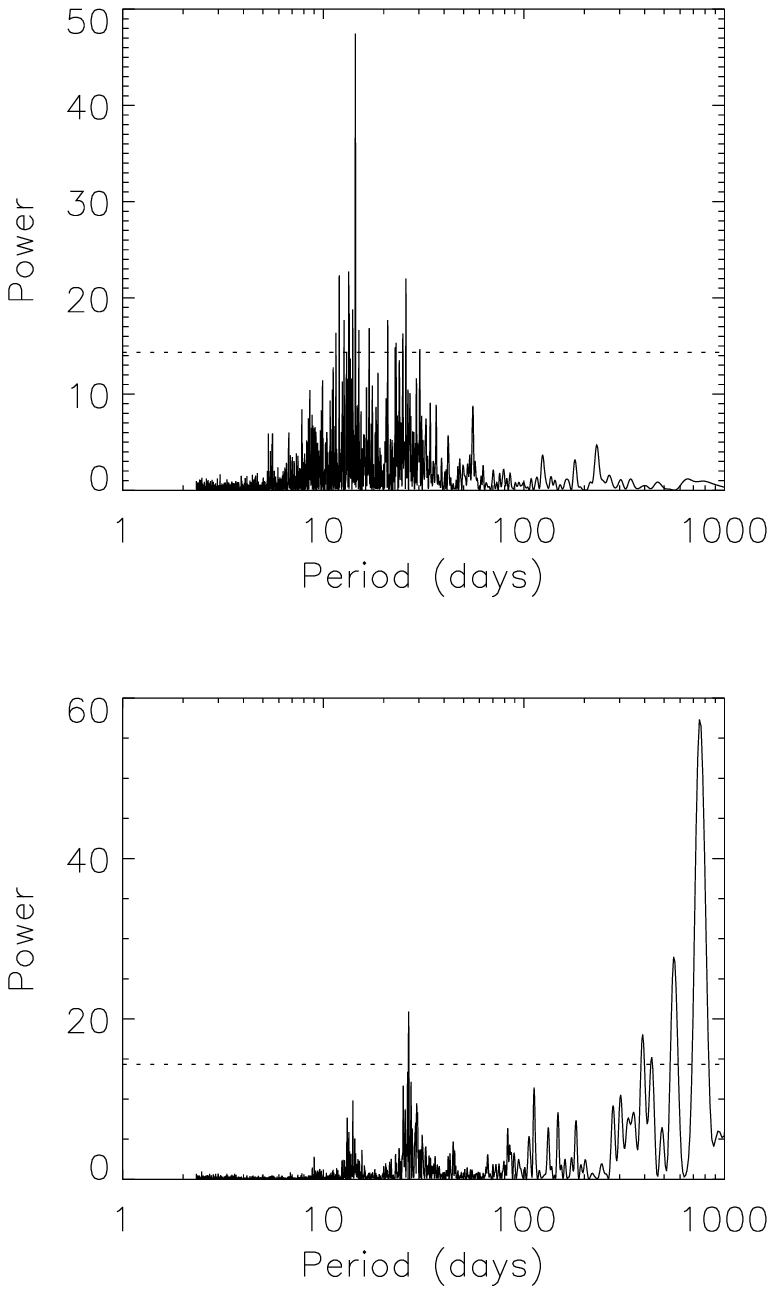}
\caption{{\it Upper panel}: Periodograms for the spot+plage RV for the whole serie. The horizontal dotted line show the level for a false-alarm probability of 1\%. {\it Lower panel}: Same for the total RV, i.e. taking the inhibition of the convective blueshift into account. }
\label{period}
\end{figure}

\begin{figure} 
\vspace{0cm}  
\includegraphics{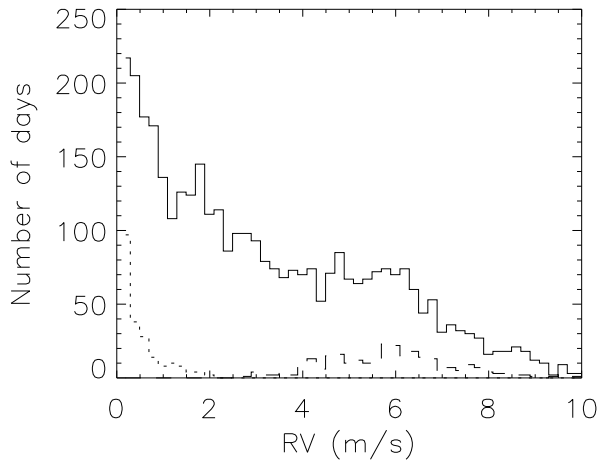}
\caption{Histogram of RV (m/s) for convection, for the whole period (solid line), low activity period (dotted line), and high activity period (dashed line).}
\label{histo_conv}
\end{figure}

\begin{table}
\begin{center}
\caption{RV rms (in m/s) for specific components.}
\label{tabrms2}
\begin{tabular}{ccccccc}
\hline \hline
 &  spots & plages & sp+pl & conv. & total \\ \hline
All & 0.34  &  0.31 &  0.33 & 2.38 & 2.40 \\
Low & 0.09  & 0.10  & 0.08  & 0.44 & 0.44 \\
High & 0.48  & 0.44  & 0.42  &  1.39 & 1.42 \\ \hline
\end{tabular}
\end{center}
\end{table}

\begin{table}
\begin{center}
\caption{RV rms and peak-to-peak amplitude (in m/s) and relative rms photometry for the three periods.}
\label{tabrms}
\begin{tabular}{cccc}
\hline \hline
 &  rms RV & ampl RV & rms phot\\ \hline
All & 2.40  & 10.8  & 3.6 10$^{-4}$\\
Low & 0.44  &  1.9 & 1.2 10$^{-4}$\\
High & 1.42  & 8.0  & 4.5 10$^{-4}$\\ \hline
\end{tabular}
\end{center}
\end{table}

\subsubsection{The spot and plage induced RV variations}

We first consider the individual time series of RV due to the photometric contributions of spots and plages. They are shown in Fig.~\ref{var_rv}. The amplitude of the spot signal is similar to what we found in Paper I (slightly smaller as expected), with maximum RV up to a few m/s during high activity periods. The RV is lower than 0.6~m/s during the minimum. The plage signal is, as expected, also higher during the high activity period (up to 1.6~m/s), i.e. about a factor two below the spot signal. The plage signal during the low activity period reaches 0.4~m/s, slightly below that of spots. It should be noted, however, that during that period the spot signal may be zero for an extended period of time, while the plage signal is always different from zero. 
Figure~\ref{varloc} (upper panel) shows the different contributions in more detail over a short period (60 days). As expected, the spot signal is partially compensated for that of plages (it was already true in the very simple computation made in Paper I), but not entirely. This can be explained as most active regions contain spots (leading to a compensation) but the area ratio between spots and plages varies \cite[e.g.][]{chapman01} and there are plages with no spot. Over the whole data set, the correlation between the spot and plage RV is indeed -0.47. 

Figure~\ref{histo} shows the distributions of RV for the spot and plage signals. The distributions are not Gaussian, as they show extended tails, and, in the case of spots, a sharp peak at zero RV (related to the significant number of days with no spot). The shape of the distribution is related to the shape of the solar cycle, i.e. to the variation in the activity level over time. A similar size distribution of structures with a constant activity level gives a Gaussian distribution. A sinusoidal shape for the solar cycle, for example, only leads to a small departure from the Gaussian shape while the proper shape gives a distribution similar to what is shown in Fig.~\ref{histo}.  
The rms RV for the whole period and the two periods corresponding to low and high activity are summarized in Table~\ref{tabrms2}. The sum of the spot and plage RV is indicated as $``$sp+pl". It confirms that at all time spots and plages provide similar rms, but as they partially compensate for each other, the resulting rms is in fact similar to that of the individual components.

Figure~\ref{period} shows the periodogram for the RV signal over the whole period for the spot+plage signal. The peaks corresponding to half the rotation period are strongly emphasized compared to the peaks at the rotation period. It may stem from the presence of active longitudes separated by 180$^{\circ}$ as discussed in Paper I. 

\subsubsection{The convection-induced RV variations}

The convection signal is much stronger than the spot and plage ones as shown in Fig.~\ref{var_rv} and Fig.~\ref{varloc}. It is also fundamentally different, as it is always positive so that all contributions add up. Therefore plages at different longitudes do not compensante for each other (as in the previous case) and the signal is maximum when the features are at the central meridian. This signal is then correlated very well with the filling factor of plages (correlation close to 1). When considering a small time series, however, as in Fig.~\ref{varloc}, even if the signal is shifted because of the convection signal, the rms is not very different from the one caused by spots and plages alone (only a factor 2 larger this time). The influence on the observed RV will therefore depend on the temporal coverage compared to the length of the cycle, as shown in  Fig.~\ref{histo_conv}. 
As previously, Table~\ref{tabrms2} shows the rms for the RV due to convection. While considering restricted periods of 9 months (low and high activity periods), the rms of the convection RV remains significantly larger than the one caused by spots and plages. 
We note a factor 6--7 between the rms due to spots and plages and the rms due to convection when computed over the whole series or at cycle minimum. The factor is twice smaller during cycle maximum. 

\subsubsection{The total RV variations}

Finally, we consider the total signal as shown in Fig.~\ref{var_rv}. The corresponding rms and maximum amplitude (peak-to-peak) are shown in Table~\ref{tabrms}, for the whole period, as well as the low and high activity periods. For the whole period, we reach an rms of 2.4~m/s, with a peak-to-peak amplitude above 11~m/s, clearly dominated by the convection signal. The amplitude is significantly lower during the low activity period, but can still reach peak-to-peak amplitudes of 2~m/s with a rms around 0.4--0.5~m/s. 
 
Figure~\ref{period} shows the periodogram for the total RV signal over the whole period. The convection RV, which dominates the total signal, emphasizes the rotation period, however, because the sign is always the same. For the same reason, the long time-scale features are also emphasized.  As the rotation period is often not known precisely, any more than the zero on the RV curves, such a difference between the two kinds of contributions may be difficult to separate.

\subsection{Comparison with observed solar RV}

There are no observations of the solar RV during the period covering our data set (cycle 23) that we are aware of. We therefore considered three previous studies covering either cycle 21 or 22. We have already compared the RV signal caused by spots in Paper I. It was not possible to use these observations to validate our simulation. By including plages it may be possible to make a more precise comparison, however.   

\cite{mcmillan93} measured the solar RV using the moon light observed in the violet part of the spectrum. They have determined an upper limit for intrinsic solar variations of $\pm$4~m/s, during cycle 22 essentially (1987--1992). This is less than our total RV variation. However, it is quite compatible with our rms RV for the spots and plages only (0.33~m/s), as well as our maximum peak-to-peak amplitude (2.4~m/s), even taking the slightly larger contrasts in the violet into account. In the wavelength domain they used, there is a very large proportion of deep lines, but as shown by \cite{gray09}, these are the lines that have the lower convective blueshift, so it is possible that their small RV comes from their being predominantly sensitive to the spot+plage RV. 

During a similar period (1983--1992), \cite{DP94} measured the solar RV in the 2.3~$\mu$m domain. They find a long-term variation with a peak-to-peak amplitude of about 30~m/s, and a large dispersion on short time scales (10--20~m/s). Considering that cycle 22 was more active than cycle 23 by about 30\%, we would expect long-term variations of about 11~m/s, i.e. 3 times less than their observation. It is quite possible that, in the wavelength domain they consider, the convective blueshift is much more pronounced, but there has been no such study that we are aware of to confirm that possibility. Their short-term variations are also significantly greater than our results for spots and plages, especially considering the reduced contrast expected at 2.3~$\mu$m.

Finally, \cite{jimenez86} measured the solar RV in the potassium line at 769.9~nm \cite[][]{brookes78} during cycle 21 (1976--1984). On long time scales, they find a peak-to-peak amplitude of 30~m/s, which is compatible with that of  \cite{DP94} and also 3 times larger than our convective signal. Their short time-scale variations can reach 20 m/s during cycle maximum, and are therefore also significantly more than our results.

\subsection{Discussion on the convection RV amplitude}

\subsubsection{Impact of the convective blueshift amplitude}

We computed the convective RV variations on a short period (60 days) when using a $\Delta V$ from 50~m/s to 300~m/s. It appears that both the RV and BVS amplitudes vary linearly with $\Delta V$. This means that it is easy to extrapolate to different convection conditions from our results. 

\subsubsection{Impact of the areas cancelling the convective blueshift}

As pointed out in Sect.~2.3.2, the reduction of the convective blueshift in the network structures outside active regions is more uncertain than in plages. We therefore recompute the convection RV while selecting structures (plages) only above 100~ppm. This corresponds to structures above 305~Mm$^2$. We recall that the convective RV is correlated very well to the filling factor, and therefore we find results that are consistent with the relative contribution of these structures over the solar cycle. As shown by \cite{Meunier03}, all structures including the smallest one (at the MDI spatial resolution) are correlated with the solar cycle, and the amplitude of variation decreases when the size decreases. Here, we find that these structures contribute to 35\% of the rms RV during the low activity period, and to 55~\% during the high activity period. 

For comparison, the same large structures contribute in a larger proportion to the plage (photometric) RV: 68~\% during the low activity period and 83~\% during the high activity period. The reason is that, in that case, because similar structures on both side of the central meridian cancel each other and because small structures are more numerous and more homogeneously distributed over the solar surface, small structures contribute less than large structures, relative to their contribution to the filling factor. For convection however, all contributions add up, so they are directly related to the filling factor. 

We also find that very small structures, for example those smaller than 80~Mm$^2$ \cite[typical network structures for][]{wang88}, should contribute to 30~\% of the filling factor at cycle maximum and 70~\% at cycle minimum. When considering structures only above that size, we find that the plage RV has almost the same rms as before (0.29~m/s instead of 0.31~m/s, confirming the insignificant contribution of the smallest strructures to the plage signal), but the rms of the convective RV is about 2~m/s instead of 2.5~m/s, so that these same small structures significantly contribute to that signal. 
As pointed out above, this only estimates the various contributions, but the exact attenuation of the convective blueshift in these regions is not well-known, even if we know that the granulation associated to them is abnormal.

Finally, the conservative threshold we have used to extract plage and network structures from MDI magnetograms implies that we are missing a significant part of the flux \cite[e.g.][]{krivova04} in very quiet regions, which may also be associated to a small attenuation of the convective blueshift, with a badly defined level. Therefore it is likely that our RV signal slightly underestimates the exact value. 

In conclusion, considering plages above 100~ppm only provides a conservative lower limit for the convection contribution, i.e. about half the total signal during cycle maximum and about a third during the minimum.

\subsubsection{Impact of the method}

To separate the photometric contribution and the RV due to the reduction of the convective blueshift in plages, we computed these two contributions separately. We also computed the RV signal on a short time series for plages that take both effects into account simultaneously; i.e., we consider a plage contribution shifted by $\Delta V$ and with the usual plage contrast in temperature. We want to compare the resulting RV: RV$_{\rm all}$ with the sum of the plage and convection RV computed before RV$_{\rm sum}$.  The difference between the two present an rms of 0.04~m/s, which is one order of magnitude smaller than the plage RV for that period (about 0.4~m/s) and of course much smaller than the convection contribution. RV$_{\rm all}$ and RV$_{\rm sum}$ are very well correlated (correlation of 0.97). There is however a systematic offset of 0.23~m/s between the two, with RV$_{\rm all}$ larger than RV$_{\rm sum}$. This value is, however, still smaller than the average RV$_{\rm all}$ (5.8~m/s). We conclude that our approach leads to the correct amplitude and variability of both contributions.

\section{Search for correlations of the radial velocity signal with other observables}

\subsection{Total solar irradiance}

\begin{figure} 
\vspace{0cm}  
\includegraphics{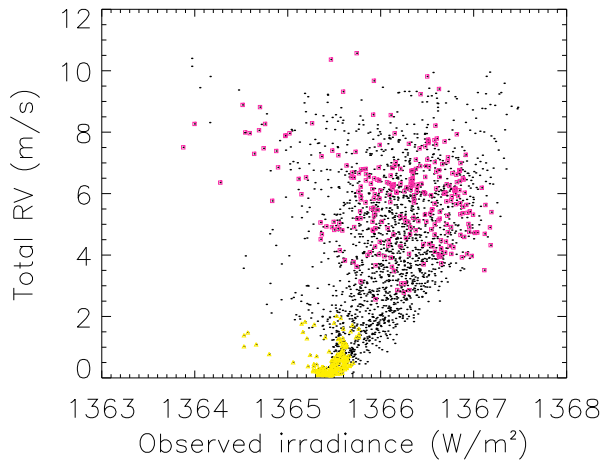}
\caption{Total RV (m/s) versus the observed TSI (W/m$^2$) for the whole data set (dots), for the low activity period (yellow triangles) and for the high activity period (pink squares). }
\label{tsi_vr}
\end{figure}

We compared the obtained RV with the observed TSI over the period 1996--2003. 
As photometry is often used as an activity criteria, it is important to check how the RV related to activity correlates with the TSI. There is no correlation between the RV due to spots and plages and the TSI. Low TSI values tend to correspond to periods with a small rms in RV, as already pointed out by \cite{kurster03}. 
When considering the total RV (Fig.~\ref{tsi_vr}), there is a small correlation between the RV and TSI over the whole period (correlation of 0.40), although this correlation tends to disappear when considering shorter periods.

\subsection{Ca index}

\begin{figure} 
\vspace{0cm}  
\includegraphics{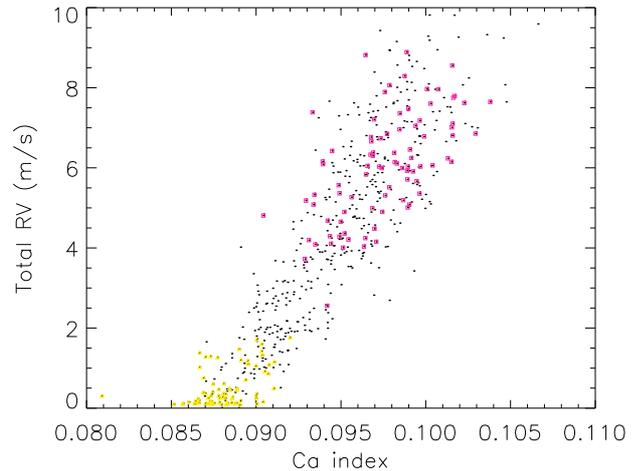}
\caption{Total RV (m/s) versus the Ca index for the whole data set (dots), for the low activity period (yellow triangles) and for the high activity period (pink squares).}
\label{ca_vr}
\end{figure}

Another commonly used indicator of stellar activity is the Ca index. It is therefore useful to study the correlation between this indicator and the observed RV. As in Paper I we use the Ca index provided by the Sacramento Peak Observatory, for the period 1996--2002.  Their is no correlation between the RV signal considering only spots and plages and the Ca index. 
However, as already noted in Paper I for spots, periods with a low Ca index correspond to a small RV dispersion, while periods with larger Ca index correspond to a larger RV dispersion. When considering the total RV, i.e. taking the convection into account, over the whole period (Fig.~\ref{ca_vr}), the correlation is strong (0.89). This is not surprising as the Ca index is related to plages, and the total RV is dominated by the convection signal (also related to plages, with no sign change).

\section{Planets}

\begin{figure} 
\vspace{0cm}  
\includegraphics{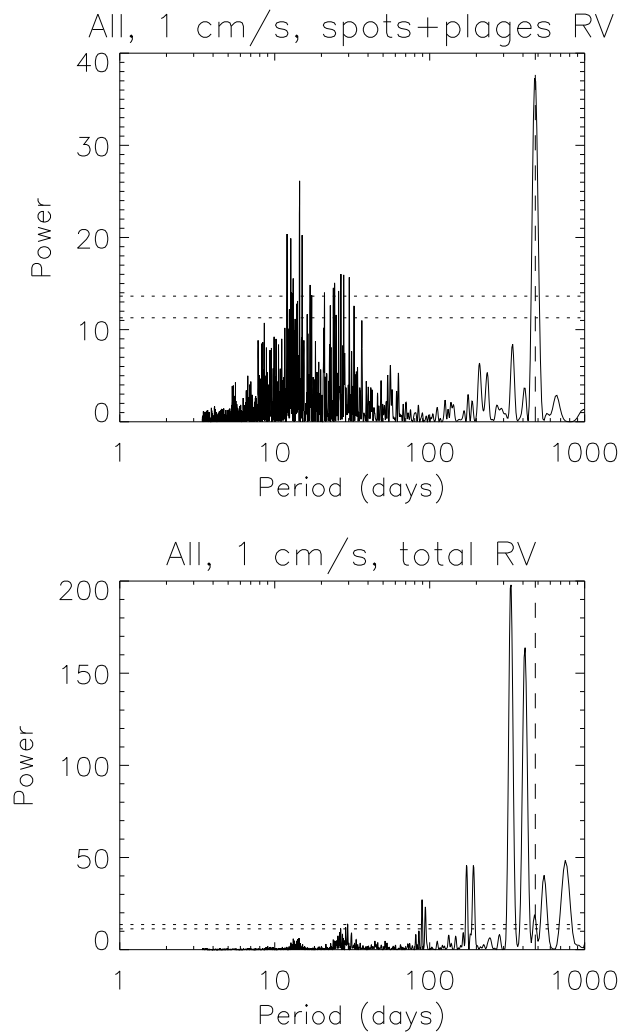}
\caption{{\it Upper panel}: Periodogram of the spot+plage RV (m/s) added to the planet (see text) and a 1~cm/s noise, over the whole period. The sampling is the original one with 4-month gaps every year. The vertical dashed line indicates the planet orbital period. The dotted lines indicate the power corresponding to false alarm probabilities of 1\% (upper line) and 10\% (lower line). {\it Lower panel}: Same for the total RV added to the planet (see text) and a 1~cm/s noise.}
\label{pla1}
\end{figure}

To test the detectability of earth-type planets in the habitable zone, we have considered, as in Paper I, a 1 M$_{\rm earth}$ planet orbiting at a 1.2~AU from the star on a circular orbit. The RV amplitude of such a planet, for a system seen edge-on, is 0.08~m/s and its orbital period 480 days.  We computed the resulting RV signal for the whole data set, as well as for two periods of low and high activity. 
The noise level is chosen to be 1~cm/s \cite[the goal for Espresso on the VLT,][]{dodorico07}, 5~cm/s, and 10~cm/s \cite[for Codex on the E-ELT,][]{dodorico07}. Different temporal samplings were considered.

\subsection{Spot and plage RV variations}

Figure~\ref{pla1} first shows the periodogram of the signal corresponding only to spots and plages, with a 1 M$_{\rm earth}$ planet orbiting at a 1.2~AU and a 1~cm/s noise level. The peak corresponding to the planet is highly significant (false alarm probability below 10$^{-12}$). The sampling is the original one with 4-month gaps every year to simulate a star that would not be visible at all times (dec$\sim-45^{\circ}$) as in Paper I. The same is true when considering only a few years of low activity (extended to cover 790 days) or high activity (extended to cover 2000 days). When degrading the temporal sampling down to 4 or 8 days, the amplitude of the planet peak decreases and becomes barely significant, especially on the shorter periods, as in Paper I. 

\subsection{Total RV variations}

When considering the total RV (including the convection contribution), again with the 1 M$_{\rm earth}$ planet orbiting at a 1.2~AU, a 1~cm/s noise level and a 4-month gap every year, Fig.~\ref{pla1} shows that the peak corresponding to the planet is not significant. It is also adjacent to larger peaks at nearby periods. This is due to the fact that convection dominates the spot+plage contribution, the noise, and the planet, so that the total RV shows a long-term variability related to the solar cycle, including on the scale of the year or a few years, hence the adjacent peaks (see for example Fig.~\ref{period}). 

If we decrease the amplitude of the convection contribution by just a factor two (see Sect.~3.3.2), the planet peak has a false alarm probabiblity of 18\%; for a factor 5 it becomes 3 10$^{-7}$. The latter is very significant, but it is still adjacent to nearby peaks that are as significant, therefore it would be very difficult to identify it as caused by a planet. Increasing the noise level or degrading the temporal sampling does not change the nature of the problem. In conclusion, we would need to decrease the convection contribution by one order of magnitude to be able to identify the planet signal, even when using the best sampling.

\section{Conclusion}

We computed the expected solar RV variations caused by the photometric contribution of spots, as well as the contributions of plages through their photometry and the suppression of the convective blueshift between 1996 and 2007. The Sun was considered to be seen edge-on and observed in conditions similar to HARPS (spectral coverage and resolution). This work complements Paper I, which only considered the spot contribution. Our simulation reproduces the observed photometry, so we are confident that their relative contributions are correct. Our approach takes the complex activity pattern observed on the Sun into account. Even if the full characterization of such a complex pattern is currently beyond reach for other stars, precise photometric observations by COROT shows that solar-like stars can exhibit several short-lived spots at the same time \cite[for example][]{Mosser09} or show a complex pattern between spots and plages \cite[e.g.][for a young solar-like star]{lanza09}. We obtain the main following results.

The photometric contributions of spots and plages only partially compensate for each other, so that the sum of these two signals is not very different from the spot signal. 
The shape of the RV distribution is not Gaussian and may provide information on the shape of the cycle, which may be a useful diagnosis tool in the future. 

The plage contribution due to the convective blueshift suppression dominates the total signal.
Unlike the previous one, the sign is always the same (a redshift when activity is present), i.e. is additive, and its long-term amplitude is about 8 m/s. The short time-scale RV rms is close to the m/s (and down to 0.4~m/s during the low activity period). 
Finally, we point out that this contribution is very line-dependent, as shown by \cite{gray09}, and therefore a precise computation must be made depending on the spectral range considered. If many lines are available, it should be possible in principle to compare the RV series computed using the strong and weak lines separately in order to identify the origin of the variation. 

The comparison with observations of the solar RV is not easy, because of different temporal coverages and wavelength domains. The photometric signal should be sensitive to the wavelength domain because the contrast (assuming a black body) is lower in the red than in the violet part of the spectrum. On the other hand, the convective blueshift is very line-sensitive. Here we have estimated the contribution when considering most lines of the visible spectrum. On the other hand, all solar RV observations have focused on a single line or a restricted part of the spectrum, for which this convective blueshift might be very different and, unfortunately, has been poorly studied.

The RV is not correlated well with the total solar irradiance. Therefore a lack of correlation between observed RV and photometry should not lead to the conclusion that the observed RV cannot be due to activity. This is true for both the photometric contribution of spots and plages to the RV and for the convection signal.   
On the other hand, when the total RV is dominated by the convection signal, we expect from the present results a strong correlation with the Ca index. It is also dependent on the variability of the Ca index. The correlation remains very small, however, when only considering the photometric contribution of spots and plages to the RV.

We also tested the influence of activity and noise on the detection of an earth-like planet at 1.2~AU on a circular orbit. When considering only the photometric contribution of the spots and plages, the results are similar to that of Paper I. A good temporal sampling is necessary to detect the planet. 
On the other hand, when also taking the convection into account, the activity-related RVs completely dominate the signal at a level of about 8~m/s, 
and it seems impossible to detect the planet. Because there are other significant peaks in the same range of periods, the convection signal would need to be at least one order of magnitude weaker to make the identification of the planet peak possible, otherwise the adjacent peaks would still be of similar amplitude.

In conclusion, the photometric contribution to RV of both spots and plages at a level comparable to the Sun should not prevent detection of earth-like planets in the habitable zone around solar-like stars, providing a very good temporal sampling (better than 8 days and over a long monitoring period) and a good signal-to-noise ratio. On the other hand, the attenuation of the convective blueshift due to active regions, at the same level, seems to make it impossible to detect such a planet in similar conditions, as the orbital period is in the same range as the long-term strong activity signal. However, the properties of this contribution are such that some diagnosis may be established more easily than with the photometric contributions.
\begin{itemize}
\item{The convective blueshift presents a strong spectral sensitivity, so that using two masks (of respectively deep and weak lines) should lead to significantly different RV variations. For example, for a $\Delta V$ of 200~m/s over all lines, we find 455~m/s for lines with depths between 0.5 and 0.95 (weak lines) and 155~m/s for lines with depths below 0.5 (strong lines), i.e. an expected factor 2.7 between two such RV series.}
\item{The correlation with the Ca index is very strong. Such a correlation could be used to correct long-term RV variations, as done by \cite{saar00}. }
\item{The BVS remains to be studied in detail, but we can expect it to be of large amplitude given the impact of that contribution on the lines, allowing it to be above the noise of future instruments. This will have to be confirmed using a realistic modelization of the line distortion when taking the convective blueshift into account. }
\end{itemize}
All these diagnosis will be very sensitive to the noise level. 

\begin{acknowledgements}
The sunspot data have been provided by USAF/NOAA. 
SOHO is a mission of international cooperation between the European Space
Agency (ESA) and NASA. 
The solar spectra were retrieved from the INSU/CNRS database BASS2000. 
The irradiance data set (version \#25) were provided by PMOD/WRC, Davos, Switzerland, and we acknowledge unpublished data from the VIRGO experiment on the cooperative ESA/NASA mission SOHO. 
We acknowledge the Sacramento Peak Observatory of the U.S. Air Force Philips Laboratory for providing the Ca index. 
We acknowledge financial support from the French Programme National de Plan\'etologie ({\small PNP, INSU}). We also acknowledge support from the French National Research Agency (ANR) through project grant NT05-4$\_$44463. 
We thank the referee, S. Solanki, for his useful comments on this paper.
\end{acknowledgements}

\bibliographystyle{aa}
\bibliography{biblio13551}

\end{document}